\documentclass[fleqn,twoside]{article}
\usepackage{espcrc2}
\usepackage{epsfig}

\usepackage{graphicx}

\newcommand{\AmS}{{\protect\the\textfont2
  A\kern-.1667em\lower.5ex\hbox{M}\kern-.125emS}}

\newcommand{\CC}{{\mathcal{C}}}
\newcommand{\dt}{{{\partial_t}}}
\newcommand{\be}{\begin{eqnarray}}
\newcommand{\ee}{\end{eqnarray}}
\newcommand{\rk}{\right)}
\newcommand{\lk}{\left(}
\newcommand{\rke}{\right]}
\newcommand{\lke}{\left[}

\title{Topological charge of Center Vortices}

\author{H. Reinhardt\address[MCSD]{Institut f\"ur Theoretische Physik \\ 
        Universit\"at T\"ubingen\\
	Auf der Morgenstelle 14 \\
	D-72076 T\"ubingen}%
        \thanks{Supported by DFG 
under grant number DFG-Re856/5-1.}}

\begin{document}

\begin{abstract}
The topological charge of center vortices is discussed in terms of the
self-intersection number of the closed vortex surfaces in 4-dimensional
Euclidian space-time and in terms of the temporal changes of the writhing number of
the time-dependent vortex loops in 3-dimensional space. 
\end{abstract}

\maketitle

\section{Introduction}
Center vortices provide an appealing picture of confinement
\cite{[Hooft]}. 
When center vortices are removed from the Yang-Mills ensemble, the
string tension \cite{[Del2]} and the quark condensate \cite{[Forc]} are lost.
Therefore one should expect that center vortices can also explain spontaneous 
breaking
of chiral symmetry. The mechanism of spontaneous breaking of chiral symmetry
seems to be tied to the topological properties of the gauge fields. My talk is
devoted to the topological charge of center vortices and is mainly based on
\cite{[4]}.

In a mathematically idealized way center vortices can be defined in a gauge
invariant way in $D$-dimensional space-time as $(D-2)$-dimensional closed
hypersurfaces (boundaries) $\partial \Sigma$ of electromagnetic flux which 
contribute a (non-trivial) center element $Z$ to the Wilson loop when 
they are non-trivially linked to the latter: 
\be
P \exp \lke -\oint\limits_{\CC} A \lk \partial \Sigma \rk \rke 
= Z ^{L \lk \CC,\partial \Sigma \rk} \quad . 
\ee  
Here $L\lk \CC, \Sigma \rk$ denotes the linking-number between the
loop $\CC$ and the hypersurfaces $\partial \Sigma$. Below I will 
consider $D = 4$. 
\section{Intersection of center vortex surfaces}
Depending on the position of the vortex surface $\partial \Sigma$ in the
4-dimensional spacetime manifold the flux of the vortex can be electric or
magnetic or both. Generically, the vortex surface $\partial \Sigma$ 
evolves in time and at a fixed time the center vortex represents a 
closed loop of magnetic flux, i.e. the $B$-field is tangential to the 
loop. There are also
purely spatial vortex surfaces existing at a single time instant only. These
vortex surface carry only electric flux being normal to the spatial surface
$\partial \Sigma$. Obviously, a non-zero topological charge
\be
\nonumber
\nu = \frac{1}{16\pi ^2} 
\int \mathrm{d}^4 x  \, \mathrm{tr} ( F \tilde{F} )
= \frac{1}{4\pi ^2}
\int \mathrm{d}^4 x  \, \vec{E} ( x ) \vec{B} (x) 
\ee  
arises when a generic vortex patch (evolving in time) and a purely spatial
vortex patch intersect\footnote{Note, by the Bianchi identity 
$\vec{\nabla} \times \vec{E} = -\dt{\vec B}$ a time-dependent 
$\vec B$-field generates an electric field $\vec E$, 
which is however, perpendicular to the $\vec B$-field.
Hence a time-dependent magnetic flux loop alone will generically 
not generate a topological charge, unless non-parallel loop segments 
intersect (see below). Note, that the dual relation 
$\vec{\nabla} \times \vec{B} = \dt{\vec E}$ (which
is part of the Yang-Mills equation of motion) is usually not 
satisfied, so that a
time-dependent $\vec E$-field not necessarily generates also 
a $\vec B$-field.} 
It is therefore not surprising that the topological charge of a center
vortex is given by its self-intersection number
$ \nu = \frac{1}{4} I \lk \partial \Sigma , \partial \Sigma \rk $.

\begin{figure}[ht]
{\epsfysize=3 cm\epsffile{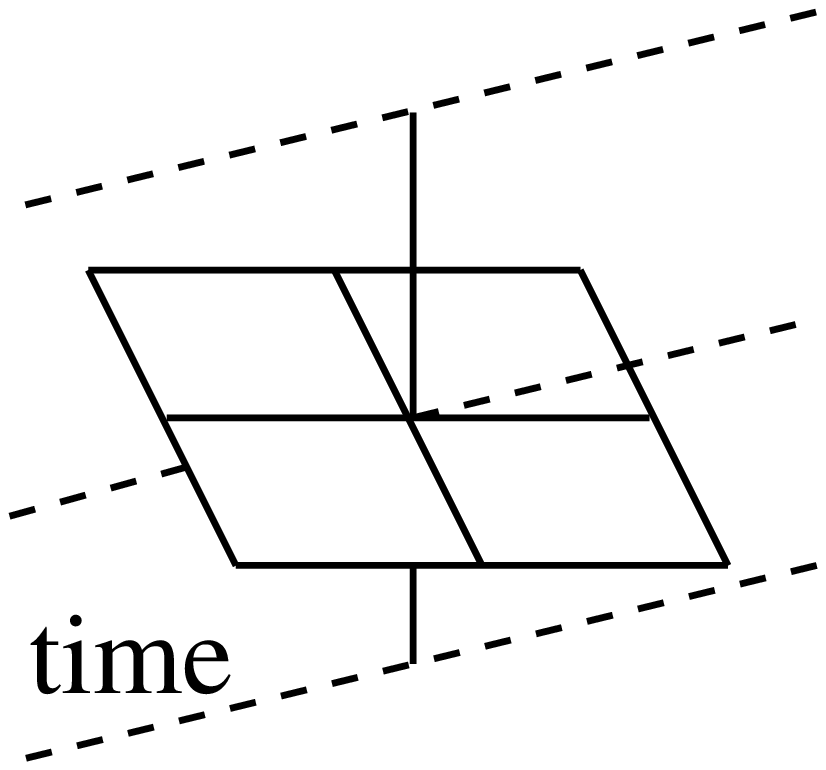}}  \hfill 
{\epsfysize=2.8 cm\epsffile{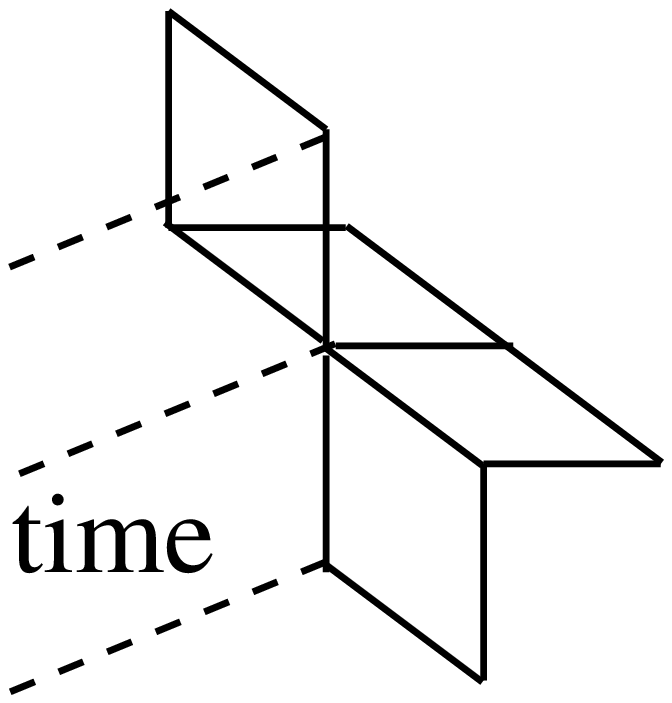}} 
\caption{\label{intersection}Intersection (left) and twisting 
(right) points. The dashed lines represent time evolution of the vortex.}
\end{figure}

The self-intersection number 
$I \lk \partial \Sigma , \partial \Sigma \rk $ receives contributions from 
two types of singular
points: (i) Transversal intersection points arising from 
the intersection of two
different surface patches (see fig.~\ref{intersection}) and (ii) 
twisting points occurring on a
single surface patch twisting around a point in such a 
way to produce 4-linearly
independent tangent vectors (see fig.~\ref{intersection}). 
Transversal intersection points yield
a contribution $\pm 2$ to the oriented intersection number,
where the sign depends on the relative orientation of the two intersecting
surface pieces. Twisting points yield always contributions of module smaller
than 2.

\begin{figure}[ht]
\centerline{
\epsfysize=5.5cm
\epsffile{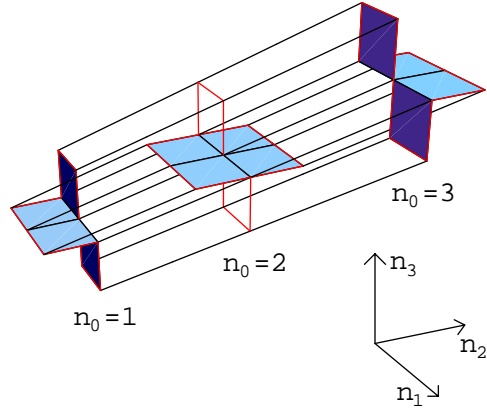}}
\caption{\label{eng}Sample lattice vortex surface configuration taken from 
\cite{[Eng2]}. At each lattice time $t=n_0 a$ (a-lattice spacing), 
shaded plaquettes are part of the vortex surface. These
plaquettes are furthermore connected to plaquettes running in time
direction.}
\end{figure}

Fig.~\ref{eng} shows a vortex with a transversal
intersection point $\lk \nu = - \frac{1}{2} \rk$ and two twisting points 
$\lk \nu = \frac{1}{8} \rk$ at the front and back, respectively, 
edges of the configuration 
at the intermediate time $(n_0 = 2)$. 
Further twisting points $\lk  \nu = \frac{1}{8} \rk$ 
occur at the initial $\lk n_0 = 1 \rk$ and final 
$\lk n_0 = 3\rk $ times,  so that the total 
topological charge of this configuration vanishes. 

\section{Writhing of center vortex loops} 

For generic center vortices representing closed magnetic flux loops 
$\CC(t)$
evolving in time the topological charge can be expressed as \cite{[4]}
\be
\nu = \frac{1}{4} \int \mathrm{d} t \, 
\partial_t W \lk \CC(t) \rk \quad , 
\ee 
where $W(\CC) = L \lk \CC, \CC \rk $ is the writhing number defined 
by the Gaussian linking number 
$L \lk \CC_1 , \CC_2 \rk$ between two loops
$\CC_1$ and $\CC_2$. 

The writhing number $W\lk \CC \rk $ is a continuous function of the 
shape of the loop $\CC$. It is not a topological invariant and thus 
not integer valued. It vanishes for planar curves or for curves 
possessing a symmetry plane, and in this sense measures the chirality 
of the loop. Furthermore, it suffers a discontinuity when
two non-parallel line-segments of the curve cross. 

\begin{figure}[ht]
\centerline{
\epsfysize=2.0cm
\epsffile{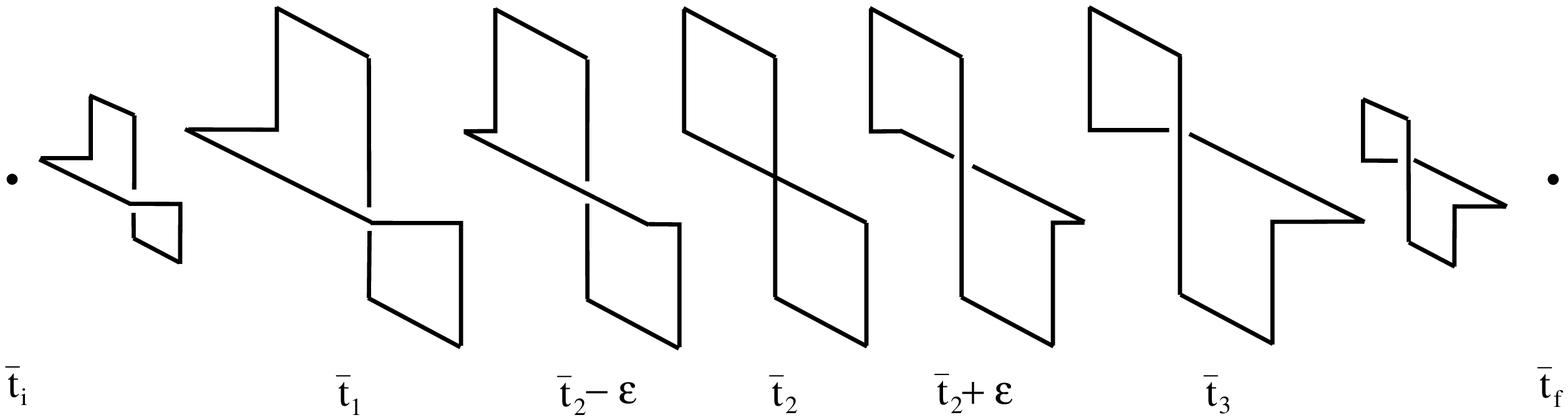}}
  \caption{\label{snap}Snap shots at characteristic time instants of the 
  continuum center   vortex loop whose lattice realization is shown in 
  fig.~\ref{eng}.}
\end{figure}

Fig.~\ref{snap} shows the time evolution of a 
closed magnetic vortex loop
in ordinary 3-space, which on a 4-dimensional lattice gives rise to the
configuration shown in fig.~\ref{eng}, after
eliminating purely spatial vortex patches, which
can be considered as lattice artifacts due to the discretization of time. 
For simplicity I have kept the cubistic form of
the vortex in $D=3$ space, so that the loops consist of straight 
line segments. 

Singular changes of the vortex loop  occur at the creation and annihilation of 
the loop (which correspond to the twisting points at $n_0 = 1, 3$ in
fig.~\ref{eng}), where the writhing number changes by 
$\Delta W  =  \frac{1}{2}$, and at the intermediate time 
$t = \bar t_2 $, where the two long line-segments cross.
This crossing changes $W$ by $(-2)$ 
(i.e.~$\Delta \nu = -\frac{1}{2}$) and corresponds in 
$D=4$ to the transversal intersection point at
$n_0 =2$, see fig.~\ref{eng}. Furthermore, when the two long loop 
segments cross at
$t=\bar t_2$ the two short horizontal loop segments at the front 
and back edges
reverse their direction, which can be
interpreted as twisting these loop segments
by an angle $\pi$. In the $D=4$ lattice realization of this vortex shown in
fig.~\ref{eng}
these twistings correspond to the twisting points at $n_0 = 2$ at the front and
back edges of the vortex. 

It turns out that transversal intersection points do not give rise to a
change of the so-called twist \cite{[4]} 
of the vortex loop, while twisting points do, which
justifies their name . 

In view of the fact that center vortices carry
spots of topological charge restricted to $|\nu| \leq \frac{1}{2}$ the recent
lattice measurement of the topological charge distribution \cite{[7]} gives
further support for the vortex picture of the QCD vacuum.

\section{Quark guides}

\begin{figure}
\centerline{\epsfxsize=7 cm\epsffile{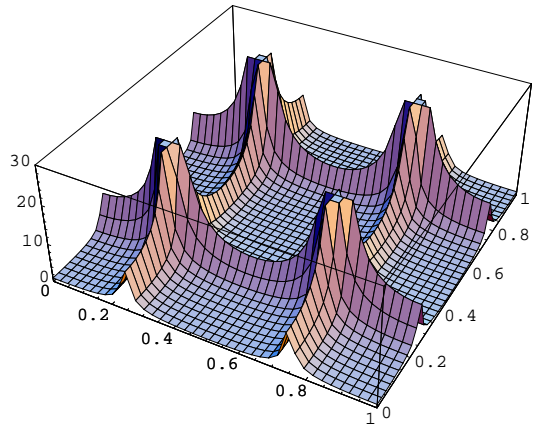}}
\caption{\label{torus}
Probability density of the quark zero modes in the background of 
four intersecting vortex sheets (with $\nu=2$) 
on the 4-dimensional torus \cite{[Tok]} shown in the subspace $x_1=x_3=0$.}
\end{figure}

A non-vanishing total topological charge requires non-oriented vortex surfaces,
which carry magnetic monopole loops at the boundary between oppositely oriented
vortex patches \cite{[4],[5]}. 
By the Athiya-Singer index theorem, $\nu = N_L - N_R$, a non-zero topological
charge $\nu$ is connected to the difference between the numbers $N_{L/R}$ of
left and right handed quark zero-modes. Fig.~\ref{torus} shows the 
probability density of the zero-modes of the quarks moving in the 
background of two pairs of intersecting center vortices on the 
4-dimensional torus \cite{[Tok]}. 
As one observes the quark zero-modes are concentrated on
the center vortex sheets and are in particular localized at the intersection
points, the spots of topological charge $\nu = \frac{1}{2}$. If the quark 
zero-modes 
dominate the quark propagator, at low energies the quarks will travel
along the center vortex sheets and can move from one vortex to an other through the
intersection points. Since the center vortices percolate in the QCD vacuum, we
expect also the percolation of the quark trajectories, which will eventually
result in a condensation of the quarks. 

\end{document}